\documentclass[prd,
twocolumn,
nofootinbib]
{revtex4}
\usepackage[usenames]{color}
\usepackage{graphicx}
\usepackage{dcolumn}
\usepackage{amsmath, amssymb, amsfonts}
\usepackage{multirow}
\usepackage{color,bm}
\usepackage[a4paper,inner=2cm,outer=1.5cm,top=1.5cm,bottom=2.0cm]{geometry}

\usepackage[colorlinks=true, urlcolor=navyblue, linkcolor=navyblue, citecolor=darkred]{hyperref}
\definecolor{navyblue}{rgb}{0, 0.0, 1.0}
\definecolor{darkred}{rgb}{1.0, 0.0, 0.0}

\newcommand\nn{\nonumber}

\newcommand{\Tr}{{\mathrm{\rm Tr}}}

\newcommand{\vecc}[1]{\mbox{\boldmath $#1$}}
\def\be{\begin{equation}}
\def\ee{\end{equation}}
\def\beq{\begin{equation}}
\def\eeq{\end{equation}}
\def\GEV{\mbox{GeV}}
\def\ba{\begin{eqnarray}}
\def\ea{\end{eqnarray}}
\def\bc{\begin{center}} 
\def\ec{\end{center}}   
\def\CG{{\mathcal G}}
\def\ba{\begin{eqnarray}}
\def\ea{\end{eqnarray}}
\def\be{\begin{equation}}
\def\ee{\end{equation}}

\begin{document}

\title{Generalized Sachs Form Factors and the Possibility of Their Measurement
\\ in Processes without and with Proton Spin Flip}

\author{M.~V.~Galynskii $^{a}$} 
\email {galynski@sosny.bas-net.by}
\author{R. E. Gerasimov $^{b,c}$}

\affiliation{$^a$ Joint Institute for Power and Nuclear Research -- Sosny, BAS, 
Minsk, 220109 Belarus \\
$^b$ Budker Institute of Nuclear Physics, 
RAS, 
Novosibirsk, 630090 Russia \\
$^c$ Novosibirsk State University, Novosibirsk, 630090 Russia}


\begin{abstract}
The differential cross section for elastic electron–proton scattering has been calculated taking into account
the two-photon exchange within the phenomenological description of the electromagnetic electron–proton
interactions. The calculation is based on the consistent evaluation of the matrix elements of the proton current
in a diagonal spin basis, which makes it possible to naturally obtain expressions for the generalized Sachs
form factors. A new method has been proposed to independently measure these form factors in the elastic
$e \vec{p} \to e \vec{p}$ process in the case where the initial proton at rest is fully
polarized along the direction of the motion of the final proton.
\end{abstract}
\maketitle


\section*{INTRODUCTION}

The study of electromagnetic proton form factors, which are important characteristics of this fundamental
particle, allows better understanding of the structure of the proton and the properties of interactions
between its constituent quarks. Experimental studies of the electric, $G_{E}$, and magnetic, $G_{M}$,
proton form factors, the so-called Sachs form factors, have been performed since the mid-1950s
\cite{Rosen,Hof58} in the elastic electron–proton scattering. In the case of unpolarized electrons
and protons, all experimental data on the behavior of the proton form factors were obtained with
the following Rosenbluth formula \cite{Rosen} for the differential cross section for the elastic
$ep\to ep$ scattering in the laboratory reference frame, where the initial proton is at rest:

\ba
\label{Ros}
\sigma=
\frac{d\sigma} {d\Omega_e}= \frac{\alpha^2E_2\cos^2(\theta_e/2)}{4E_1^{\,3}\sin^4(\theta_e/2)}
\frac{1}{1+\tau_p} \left(G_E^{\,2} +\frac{\tau_p}{\varepsilon}G_M ^{\,2}\right).
\ea
Here, $\tau_p=Q^2/4M^2$, where $Q^2=
4E_1 E_2\sin^2(\theta_e/2)$
is the square of the momentum transferred to the proton and M is the proton mass;
$E_1$, $E_2$, and $\theta_e$ are the energies of the initial and final electrons
and the angle of scattering of the electron in the laboratory reference frame, respectively;
$\alpha=1/137$ is the fine structure constant; and $\varepsilon=[1+2(1+\tau_p)\tan^2(\theta_e/2)]^{-1}$
is the degree of the linear (transverse) polarization of the virtual photon
\cite{Dombey,Rekalo74,AR,GL97} varying in the range $0 \leqslant \varepsilon \leqslant 1$.
The Rosenbluth formula (\ref{Ros}) is obtained in the one-photon exchange approximation
with zero mass of the electron.

According to the Rosenbluth formula, the main contribution to the $ep \to ep$ cross section at high
$Q^{\,2}$ values comes from the term proportional to $G_M^{\,2}$, which
reduces the accuracy of the separation of the $G_E^{\,2}$ contribution.
For this reason, the use of the Rosenbluth formula to experimentally determine the form factors
$G_E$ and $G_M$ gives significant uncertainties at $Q^{\,2}\geqslant 1$ GeV$^2$.
The method of measurement of Sachs form factors based on Eq. (\ref{Ros}) is called the Rosenbluth technique.
The experimental dependence of $G_E$ and $G_M$  on $Q^{\,2}$ was determined
using the Rosenbluth technique. Up to $Q^2\approx 6$ GeV$^2$, this dependence is well described
by the dipole approximation

\ba
G_E \approx G_M/\mu_p \approx G_D(Q^2) = (1+Q^2/\,0.71)^{-2}\,,
\label{eq:GMSL}
\ea
where $\mu_p=2.79$ is the magnetic moment of the proton.
In Eqs. (\ref{eq:GMSL}) and (\ref{linfit}), $Q^2$ is given in units of GeV$^2$.
The ratio $R \equiv \mu_p G_E/G_M$ is $R \approx 1$.

Akhiezer and Rekalo \cite{Rekalo74} proposed a method for measuring the ratio of the Sachs
form factors based on polarization transfer from the longitudinally polarized initial
electron to the final proton. This method involves the following expression obtained
in \cite{Rekalo74} for the ratio of the form factors $G_E$ and $G_M$ in terms of the ratio
of the transverse, $P_t$, and longitudinal, $P_l$, polarizations of the scattered proton:
\ba
\label{AxRek}
R \equiv  \frac{\mu_p\, G_E}{G_M}=-\frac{P_t}{P_l}\frac{E_1+E_2}{2M} \tan\left(\frac{\theta_e}{2}\right).
\ea

High-precision experiments based on Eq. (\ref{AxRek}) were performed at the Thomas Jefferson National Accelerator
Facility (JLab, United States) \cite{Jones00,Gay01,Gay02,Pun05,Puckett10,Puckett12} for the
range $0.5\, \GEV^2 \leqslant Q^2   \leqslant 8.5\, \GEV^2$. It appeared that the
ratio $R$ decreases linearly with increasing $Q^2$ in the
range $0.5\, \GEV^2 \leqslant Q^2  \leqslant 5.6\, \GEV^2$ as
\begin{equation}
R =1-0.13\,(Q^2-0.04)\,
\approx 1-Q^2/8\, ,
\label{linfit}
\end{equation}
which indicates that $G_E$ decreases more rapidly than $G_M$.

Repeated, more accurate measurements of the ratio $\mu G_E/G_M$ \cite{Puckett10,Puckett12,Qattan}
only confirmed discrepancy with results obtained using the Rosenbluth technique.
The current status of this problem was reviewed in detail in \cite{Perdrisat2007,ETG15}.

To resolve the appearing contradiction, it was assumed that discrepancy in experiments could appear
because the contribution of two-photon exchange was disregarded in the corresponding analysis,
which initiated numerous theoretical
\cite{Guichon2003,Brodsky2005,Tjon2007,Borisyuk2007,Borisyuk2008,Borisyuk2011,Kivel2013} and experimental
works \cite{Meziane11,Puckett17} (see also reviews \cite{Carlson2007,Arrington2011} and references
therein).

The contribution of two-photon exchange to the $ep \to ep$ cross section for elastic scattering has already
been measured in three experiments. These are the experiment at the VEPP-3 storage ring in Novosibirsk
\cite{Gramolin2015}, the EG5 CLAS experiment (JLab) \cite{CLAS2015}, and the
OLYMPUS experiment at the DORIS accelerator (DESY, Germany) \cite{OLYMPUS2017}.
The preliminary results of works \cite{Gramolin2015,CLAS2015,OLYMPUS2017} show that the inclusion
of the contribution of two-photon exchange, as should be expected,
can remove contradictions to $Q^2 \leqslant 2$ -- 3 GeV$^2$.

In view of significant discrepancies between the measurements of the ratio of the Sachs form factors by
two experimental methods, it would be very important to perform measurements by other independent methods.
A new method for the experimental measurement of squares of the Sachs form factors was proposed in
\cite{JETPL18} in the one-photon approximation. In this method, the form factors $G_E^{\,2}$ and
$G_M^{\,2}$ can be determined independently from each other from direct
measurements of cross sections for the elastic $e \vec{p} \to e\vec{p}$
process without and with proton spin flip in the case where the initial proton
at rest is fully polarized along the direction of motion of the scattered proton.
The aim of this work is to demonstrate that the method proposed in \cite{JETPL18}
is also applicable in the two-photon approximation and makes it possible to similarly measure
the squares of absolute values of the generalized Sachs form factors
$|{\mathcal G}_E|^{\,2}$ and $|{\mathcal G}_M|^{\,2}$.  The proposed method appears to be applicable
because the polarization structure of the cross sections for the $e \vec{p} \to e\vec{p}$
process is the same in the one- and two-photon approximations. The calculation of the cross section
in the two-photon approximation is performed using the method of calculation of the matrix elements of
quantum electrodynamics (QED) processes in the diagonal spin basis \cite{Sik84}, where the little Lorentz group
\cite{FIF70,GL} common for two particles with different momenta is implemented. The diagonal spin basis
allows naturally determining the generalized Sachs form factors at the stage of calculation of matrix elements
of the proton current. In this case, the cross section for the $e \vec{p} \to e\vec{p}$ process
in an arbitrary reference frame contains only $|{\mathcal G}_E|^{\,2}$ and $|{\mathcal G}_M|^{\,2}$, i.e.,
"is diagonalized" in the language used in \cite{Borisyuk2008}.


\section*{DIAGONAL SPIN BASIS}

In the diagonal spin basis, spin 4-vectors $s_{1}$ and $s_{2}$ of the protons
with the 4-momenta $q_{1}$ and $q_{2}$ ($s_{1} q_{1} = s_{2} q_{2} = 0$,
$s_{1} ^{2} = s_{2} ^{2} = - 1, q_1^2=q_2^2=M^2$) or, correspondingly,
with the 4-velocities $v_{1}=q_1/M$ and $v_{2}=q_2/M$ have the form \cite{Sik84}
\ba
\label {DSB}
s_{1} = - \; \frac { (v_{1} v_{2}) v_{1} - v_{2}} {\sqrt{(v_{1}v_{2} )^{2} - 1 }} \; , \; \;
s_{2} =  \frac { ( v_{1} v_{2})v_{2} - v_{1}} {\sqrt{ ( v_{1}v_{2} )^{2} - 1 }} \;.
\ea
Spin 4-vectors (\ref{DSB}) obviously do not change under transformations
of the little Lorentz group common for particles with 4-momenta $q_1$ and $q_2$.

We consider spin vectors of the diagonal spin basis (\ref{DSB}) in the laboratory
reference frame. In the general case, the spin 4-vector $s$ of a spin-1/2 particle with the
4-momentum $q$ has the form
\ba
s=(s_{0}, \vecc s), \; s_{0}=\vecc v  \vecc c, \; \vecc s =\vecc c
+ \frac{\;(\vecc c \,\vecc v)\vecc v\;}{1+v_{0}}\;,
\label{spinv}
\ea
where $\vecc c$ is an arbitrarily directed three-dimensional
unit ($\vecc c^2=1$) vector called the spin projection axis.

In the laboratory reference frame, where $q_1=(M,\vecc 0)$ and $q_2=(q_{20}, \vecc q_2)$,
the spin 4-vectors $s_{1}$ and $s_{2}$ of the diagonal spin basis (\ref{DSB}) have the form
\ba
\label{DSB_LSO1}
s_1=(0,\vecc n_2 )\,, \; s_2= (|\vecc v_2|, v_{20}\, \vecc {n_2})\,, \,\vecc n_2=  \vecc {q_2}/|\vecc q_2|\,.
\ea
This means that the spin projection axes $\vecc c_{1}$ and $\vecc c_{2}$ for
the initial and final particles in the laboratory reference frame coincide with
the direction of motion of the final particle
\ba
\vecc c_{1} =\vecc c_{2}=\vecc n_2=  \vecc {q_2}/|\vecc q_2|\,.
\label{LSO}
\ea

For the system of two particles with different momenta $q_1=(q_{10}, \vecc {q_1})$
(before interaction) and $q_2=(q_{20}, \vecc {q_2})$ (after interaction),
the possibility of the simultaneous projection of the spins on a single common
direction in an arbitrary reference frame is determined
by the three-dimensional vector \cite{FIF70}
\ba
\vecc a = \vecc q_{2}/q_{20} - \vecc q_{1}/q_{10}\,.
\label{os}
\ea
This result was obtained within the vector parameterization of the little Lorentz
group $L_{q_1,q_2}$, common for two particles with 4-momenta $q_1$ and $q_2$
($L_{q_1 q_2} q_1 =q_1, L_{q_1 q_2} q_2 =q_2$) \cite{FIF70,GL}.

The term "diagonal spin basis" is introduced because the three-dimensional vector
$\vecc a$ given by Eq. (\ref{os}) is the difference of two vectors and is geometrically
the diagonal of a parallelogram.

The coincidence of the little Lorentz groups for particles with the 4-momenta $q_1$
and $q_2$ in the diagonal spin basis specified by Eqs. (\ref{DSB}) is responsible for a
number of remarkable properties of this basis. In particular, in the diagonal spin basis
given by Eqs. (\ref{DSB}), the spin projection operators $\sigma_{1}$ and $\sigma_{2}$,
as well as the raising and lowering spin operators $\sigma_{1}^{\pm\delta}$
and $\sigma_{2}^{\pm\delta}$, for the initial and final particles coincide with each other and
have the form \cite{GS89,GS98}:
\ba
&&\sigma = \sigma_{1} = \sigma_{2} =\gamma^{5} {\hat s_1} {\hat v_1} =
\gamma^{5} {\hat s_2} {\hat v_2}  = \gamma^{5} {\hat b}_{0} {\hat b}_{3} ,
 \label{spop6a} \\
&&\sigma^{\pm\delta} = \sigma_{1}^{\pm\delta} =\sigma_{2}^{\pm\delta}
= - 1/2 \,\gamma^{5}\, \hat{b}_{\pm\delta}\,,
 \label{spop6b}\\
&&\sigma u^{\delta}(q_{i}) = \delta u^{\delta}(q_{i}), \sigma^{\pm\delta} u^{\mp\delta}(q_{i})
= u^{\pm\delta}(q_{i}).
\label{spop6c} 
\ea
Here, $u^{\delta}(q_{i})$=$u^{\delta}(q_{i}, s_{i})$ are the bispinors of states of the
particles $(i=1,2)$; $\hat a = a_{\mu} \gamma^{\mu}$ is an arbitrary Dirac operator;
$\gamma^{\mu}$ and $\gamma^5$ are the Dirac matrices; and $b_{\pm\delta} = b_{1} \pm i\, \delta b_{2}$
are the circular 4-vectors, where $\delta = \pm 1$, $b_{\pm\delta}^{~2}=0$,
$b_{\pm\delta} b_{\mp\delta}=-2$.

In Eqs. (\ref{spop6a}) and (\ref{spop6b}), to construct the spin operators,
the following tetrad of orthonormalized 4-vectors $b_{A}$ $(A = 0, 1, 2, 3)$ is used:
\ba
\label{OBV}
&&b_0=q_+/\sqrt{q_+^2}\; , \; b_{3} = q_-/ \sqrt{-q_- ^2} \; \;, \\
&& (b_1)_{ \mu} = \varepsilon_{\mu \nu \kappa \sigma}b_0^{\nu}b_3^{\kappa}b_2^{\sigma},\;
(b_{2})_{\mu} = \varepsilon_{\mu \nu \kappa \sigma}q_1^{\nu}q_2^{\kappa} r^{\sigma}/\rho ,\nn
\ea
where $q_+=q_2+q_1$, $q_-=q_2-q_1$, $\varepsilon_{\mu\nu\kappa\sigma}$ is the Levi-
Civita tensor  ($\varepsilon_{1230}=1$), $r$ is the 4-momentum of a
particle involved in the reaction different from $q_{1}$ and $q_{2}$
, and $\rho$ is determined from the normalization conditions
$ b_{1}^{2} = b_{2}^{2} = b_{3}^{2}=-b_{0}^{2}=-1$. The coincidence of the
spin operators in the diagonal spin basis (\ref{DSB}) makes it
possible to separate in the covariant form the interactions
without and with spin flip of particles involved in
the reaction and, thereby, to trace the dynamics of the spin interaction.

\section*{METHOD OF THE CALCULATION OF THE MATRIX ELEMENTS OF QED
PROCESSES IN THE DIAGONAL SPIN BASIS}

The amplitudes of the QED processes in the scattering channel have the form
\beq
M^{\pm\delta,\delta} = \overline {u}^{\pm \delta}(q_{2}) Q_{in} u^{\delta}(q_{1}) ,
\label{MQED}
\eeq
where $u^{\delta}(q_{i})$=$u^{\delta}(q_{i},s_i)$ are the bispinors of the initial
and final states of fermions satisfying the normalization
condition $\overline {u}^{\delta}(q_{i})\, u^{\delta}(q_{i}) = 2M$,
where $q_{i}^{2} = M^{2} \, (i = 1, 2)$, and $Q_{in}$ is the interaction operator.

The matrix elements given by Eq. (\ref{MQED}) can be reduced to the trace
of the product of the operator:
\ba
\label{MQED1}
  M^{\pm\delta,\delta} = \Tr (P_{21}^{\pm\delta,\delta} Q_{in} ), \\
  P_{21}^ {\pm\delta,\delta} = u^{\delta}(q_{1})  \overline {u}^{\pm \delta}(q_{2}).
\label{P21pma}
\ea

The operators $P_{21}^{\pm \delta,\delta}$ given by Eq. (\ref{P21pma}) can be determined
by a number of methods \cite{GS98,Cedrik18}. In the approach \cite{GS98} used in this work,
in contrast to, e.g., \cite{Cedrik18}, the determination of $P_{21}^{\pm \delta,\delta}$
is reduced to finding the operators $T_{21}$ and $T_{12}$ such that
\ba
u^{\delta}(q_{2})=T_{21}u^{\delta}(q_{1}) , \, u^{\delta}(q_{1})=T_{12}u^{\delta}(q_{2}),
\ea
which have the properties $T_{12}=T_{21}^{-1}$ and $T_{21}T_{12}=1$.

As a result, the operator $P_{21}^{\delta,\delta}$ specified by Eq. (\ref{P21pma}) is
obtained in the form
\ba
P_{21}^{\delta,\delta} = u^{\delta}(q_{1}) \, \overline {u}^{\delta}(q_{1}) \,T_{12}
= \tau_{1}^{\delta}\, T_{12}=T_{12} \, \tau_{2}^{\delta}\,,
\label{P21pp}
\ea
where
\ba
\tau^{\delta}_i =u^{\delta}(q_{i}) \overline {u}^{\delta}(q_{i})=
1/2\, (\,\hat{q}_{i} + M) (\, 1 -  \delta \gamma^{5} \hat{s}_{i} )
\label{po1}
\ea
are the projective operators of states of particles with
4-momenta $q_i$ and spin 4-vectors $s_i$ ($q_is_i=0, s_i^2=-1$, $i=1,2$).
The operator $P_{21}^{-\delta,\delta}$ given by Eq. (\ref{P21pma}) is the
product of the operators $\sigma^{+\delta}$ (\ref{spop6b}) and $P_{21}^{-\delta,-\delta}$ (\ref{P21pp}):
\ba
P_{21}^{-\delta,\delta} =\sigma^{+\delta}  P_{21}^{-\delta,-\delta}
= \sigma^{+\delta} \tau^{-\delta}_1 T_{12}\,  
 = \sigma^{+\delta} T_{12} \tau^{-\delta}_2.
\label{P21pm}
\ea

In the diagonal spin basis (\ref{DSB}), $T_{21}$ and $T_{12}$ coincide
with each other and have the form \cite{GS98}
\ba
T_{21} = T_{12} = \hat b_{0} .
\label{T120} \nn
\ea
As a result, the operators $P_{21}^{\pm\delta,\delta}$ (\ref{P21pma}) are obtained in the
form \cite{GS98}
\ba
\label{P21pp2}
P_{21}^{\delta,\delta} = ( \hat q_{1} + M ) \, \hat b_{\delta} \,  \hat b_{0}
\; \hat b_{\delta}^{\ast} /4 \, ,  \\
P_{21}^{-\delta,\delta} = \delta (\hat q_{1} + M) \; \hat b_{\delta} \; \hat b_{3} /2\,,
\label{P21pm2}
\ea
where $b_{\delta}^{\ast} =b_{-\delta}  = b_{1} - i \delta b_{2},  b_{\delta} b_{\delta}^{\ast}=-2,
b_{\delta}^2= b_{\delta}^{\ast 2}=0$.

The general structure of the dependence of the squares of absolute values
of the matrix elements (\ref{MQED}) on the polarization of particles can be established in
some cases from their general form using the symmetries of electromagnetic interactions.
To demonstrate this, we rewrite the matrix element (\ref{MQED}) in the most
general form
\beq
M(\delta_1,\delta_2) \equiv  M^{\pm\delta_2,\delta_1}
= \overline {u}^{\pm \delta_2}(q_{2}) Q_{in} u^{\delta_1}(q_{1})\,.
\label{MQED2}
\eeq
We introduce the polarization factors
\ba
\omega_{+}=(1 + \delta_1 \delta_2)/2, \, \,\omega_{-}=(1 -\delta_1 \delta_2)/2\,.
\label{omegi}
\ea
At $\delta_{1,2}=\pm1$, they have the properties
\ba
\omega_{\pm}^2=\omega_{\pm},\; \omega_{\pm}\omega_{\mp}=0.
\label{prop_om}
\ea
The matrix element given by Eq. (\ref{MQED2}) satisfies the relation
\ba
M(\delta_1,\delta_2)= \omega_+  M^{+\delta_2,\delta_1} + \omega_-  M^{-\delta_2,\delta_1} .
\label{MQED3}
\ea
In view of the properties of polarization factors $\omega_{+}$ and $\omega_{-}$
specified by Eqs. (\ref{prop_om}) at $\delta_{1,2}=\pm1$, we have
\ba
|M(\delta_1,\delta_2)|^2&=& \omega_+ | M^{+\delta_2,\delta_1}|^2 + \omega_-  |M^{-\delta_2,\delta_1}|^2 .
\label{MQED4}
\ea
Because of the conservation of spatial parity, spin correlations in Eq. (\ref{MQED4})
for the process $e \vec p \to e \vec p$, where electrons are unpolarized, should be absent except for
those contained in $\omega_{+}$ and $\omega_{-}$. This means that $|M^{\pm\delta_2,\delta_1}|^2$
are independent of $\delta_1$ and $\delta_2$, and the square of the absolute value
of the matrix element given by Eq. (\ref{MQED2}) averaged and summed over polarizations has the form
\beq
\overline{|M(\delta_1,\delta_2)|^2}= | M^{\uparrow\uparrow}|^2 +  |M^{\downarrow\uparrow}|^2 .
\label{MQED5}
\eeq

\section*{MATRIX ELEMENTS OF PROTON CURRENT IN THE ONE-PHOTON APPROXIMATION}

The matrix elements of the elastic process $e \vec p \to e \vec p$
\ba
e(p_1)+p(q_1,s_1) \to e(p_{2}) + p(q_2,s_2)
\label{EPEP}
\ea
is the product of the electron and proton currents:
\ba
\label{Mepep}
&& M_{ep\to ep} = 4\pi \alpha T / q^2\,, \\
&& T\equiv T^{\pm\delta,\delta }=(J_{e})^{\mu} ( J^{\pm \delta,\delta }_{p} )_{\mu} .
\label{T2pm}
\ea
In the one-photon exchange approximation, the currents $(J_{e})_{\mu}$
and $(J_p)^{\mu}$ have the form
\ba
\label{Je}
 (J_{e})^{\mu} &=& \overline{u}(p_{2}) \gamma^{\mu} u(p_{1}) \,,  \\
\label{Jp}
 (J_p)_{\mu} &=&\overline{u}(q_2) \Gamma_{\mu}^{1\gamma}(q^{2}) u(q_1) \,, \\
\Gamma_{\mu}^{1\gamma}(q^{2}) &=& F_{1} \gamma_{\mu} + \frac{F_{2}} {4M}
( \; \hat q \gamma_{\mu} - \gamma_{\mu} \hat q ) \, .
\label{Gamuepep}
\ea
Here $u(p_{i})$ and $u(q_{i})$ are the bispinors of electrons and protons
with 4-momenta $p_{i}$ è $q_{i}$, respectively, where
$p_{i}^{2} = m^{2}$ and $q_{i}^{2} = M^{2}$, having the properties
$\overline{u}(p_{i})u(p_{i})=2m$ and $\overline{u}(q_{i}) u(q_{i})= 2M$ $(i = 1,2)$;
$F_{1}$ and $F_{2}$ are the Dirac and Pauli proton form factors, respectively;
$q = q_-=q_{2}-q_{1}$ is the 4-momentum transferred to the proton; and $s_1$ and $s_2$
are the polarization 4-vectors of the initial and final protons, respectively.

The cross section for the process $e \vec p \to e \vec p$ has the form
\ba
\label{Modepep}
\frac {d \sigma}{d |t|}= \frac{ \pi \alpha^2 }{4 I^2 }\, \frac {|T|^2}{q^4}  ,
\ea
where $ I^2=(p_1q_1)^2-m^2M^2$ and $|t|=Q^2=-q^2$.

The matrix elements of the proton current (\ref{Jp}) calculated in the diagonal spin
basis (\ref{DSB}) using Eqs. (\ref{MQED1}), (\ref{P21pp2}), and (\ref{P21pm2})
have the form \cite{Sik84,GS98}
\ba
\label {Jepep-pp0}
( J^{\delta,\delta }_{p} )_{\mu} = 2 M \,G_{E} ( b_{0} )_{\mu} \, , ~~~~~~\\
( J^{-\delta,\delta }_{p} )_{\mu}= - 2 \,\delta M \, \sqrt{\tau_p} \,G_{M} (b_{\delta })_{\mu}\, ,
\label {Jepep-pm0}
\ea
where
\ba
G_{E} = F_{1} -\tau_p \, F_{2} \, , \; G_{M} = F_{1} + F_{2}
\label {FFSep}
\ea
are the Sachs form factors.
Consequently, the Sachs form factors $G_E$ and $G_M$ in the matrix elements of the
proton current corresponding to transitions without, Eq. (\ref{Jepep-pp0}),
and with, Eq. (\ref{Jepep-pm0}), proton spin flip in the diagonal spin basis
are factorized. Because of this property, these form factors have a fundamental physical
meaning as quantities determining the probabilities of transitions of the proton
without and with spin flip in the case where the axes of the spin projections
coincide and have the form of Eq. (\ref{os}).

In the case of unpolarized electrons, the matrix elements of the proton current
$J_p^{\pm \delta,\delta}$ (\ref{Jp}) reduce the squares of the absolute values of the amplitudes
$|T^{\pm \delta,\delta}|^2$ (\ref{T2pm}) to the trace of the product of operators:
\ba
|T^{\pm \delta,\delta}|^2=2 \cdot \Tr(\tau_{2}^e \gamma_\mu \tau_{1}^e \gamma_\nu)
(J_p^{\pm \delta,\delta})^\mu (J_p^{\pm \delta,\delta \ast})^\nu \, .
\label{Tquad1}
\ea
Here, the asterisk $\ast$ means complex conjugation and
\ba
\tau_{i}^e = 1/2\, (\,\hat{p}_{i} + m)
\label{poe}
\ea
are the projective operators of the electron ($i=1, 2$).

According to Eq. (\ref{MQED4}), the quantity $|T|^2$ determining
the cross section (\ref{Modepep}) for the $e \vec p \to e \vec  p $ process has the form
\ba
|T_{\delta_1, \delta_2}|^2 =\omega_+ |T^{+ \delta,\,\delta}|^2
+\omega_- |T^{- \delta,\,\delta}|^2\,.
\label{Td2}
\ea

In the case of unpolarized particles, the spin-averaged square of the absolute
value of the amplitude $\overline{|T|^2}$ has the form
\ba
\overline{|T|^2} =|T^{+ \delta,\,\delta}|^2 + |T^{- \delta,\,\delta}|^2\,.
\label{Td3}
\ea

\bc
{\bf
CROSS SECTION FOR THE $e \vec p \to e \vec p$ PROCESS
IN THE ONE-PHOTON APPROXIMATION}
\ec

In terms of the matrix elements of Eqs. (\ref{Jepep-pp0}) and
(\ref{Jepep-pm0}), the calculation of $|T|^2$ (\ref{T2pm}) is reduced to the calculation
of traces
\ba
|T^{+ \delta,\,\delta}|^2& =&4M^2 \, G^{\,2}_E
\, \Tr ((\hat p_2 +m)\hat b_0(\hat p_1 +m)\hat b_0) /2\,, \nn \\
|T^{ -\delta,\,\delta}|^2&=& 4M^2 \tau_p G^{\,2}_M
\Tr ((\hat p_2 +m)\hat b_{\delta}(\hat p_1 +m) \hat b^*_{\delta})/ 2\,.\nn
\ea
The simple calculations give
\ba
\label{Wep pp}
|T^{+ \delta,\,\delta}|^2&=& \frac{G^{\,2}_E}{1+\tau_p}\,  Y_1 ,\,
|T^{- \delta,\,\delta}|^2 =\frac{\, \tau_p \, G^{\,2}_M} {1+\tau_p} \, Y_2\,,~~~\\
Y_1&=&(p_+q_+)^2+q_+^2q_-^2\,,\\
Y_{2}&=&(p_+ q_+)^2-q_+^2(q_-^2+4 m^2)\,,
\label{Wep pm}
\ea
where $p_+=p_1+p_2$, $q_+=q_1+q_2$, and $q_-=q_2-q_1=q$.

We note that the quantities $|T^{\pm \delta,\,\delta}|^2$ (\ref{Wep pp}) are independent
of the polarizations of protons, as mentioned above.
As a result, the differential cross section for the
$e \vec{p} \to e \vec{p}$ process in the diagonal spin basis in an arbitrary
reference frame is given by the expression
\ba
\label{crepep1}
\frac {d \sigma_{\delta_1, \delta_2}}{d |t|}= \frac{ \pi \alpha^2 }{4 I^2 (1+\tau_p)}
\left(  \omega_+ G_{E}^{2}  Y_{1} + \omega_- \tau_p G_{M}^{2}  Y_{2}  \right) \frac{1}{q^4}.~~
\ea

Substituting $\delta_2=0$ into Eq. (\ref{crepep1}) and doubling the result, we obtain
the cross section for the $ep \to ep$ process, where all particles are unpolarized,
in an arbitrary reference frame in the form
\ba
\label{crepep}
\frac {d \sigma}{d |t|} = \frac{ \pi \alpha^2 }{4 I^2\, (1+\tau_p)}  \, ( \, G_{E}^{2} \; Y_{1}
+ \tau_p \; G_{M}^{2} \; Y_{2} \, )\, \frac{1}{q^{4}} \, .
\ea
This expression coincides with Eq. (34.3.3) in \cite{AB}.

It is convenient to represent the expression for $Y_1$
and $Y_2$ in terms of the Mandelstam variables
\ba
s=(p_1+q_1)^2, \, t=(q_2-q_1)^2, \, u=(q_2-p_1)^2. \nn
\ea
The inversion of the relation gives
\ba
p_+q_+=s-u, \, q_+^2=4M^2-t, \, q_-^2=t, \, \tau_p=-t/4M^2, \nn
\ea
from where
\ba
&& Y_1=(s-u)^2+(4M^2-t)\,t, \; \;\\
&& Y_2=(s-u)^2-(4M^2-t)(t+4m^2)\,.
\ea

The quantity $4I^2$ appearing in Eq. (\ref{Modepep}) can be expressed in terms
of $s$, $t$, and $u$ in the form
\ba
4I^2=(s-(M+m)^2)(s-(M-m)^2)=\lambda(s,m^2,M^2), \nn
\ea
where $\lambda(s,m^2,M^2)$ is the K\"{a}ll\'{e}n  function.

The cross section given by Eq. (\ref{crepep}) expressed in
terms of the variables $s$, $t$, and $u$ coincides with
Eq. (139.4) in \cite{BLP}.

We introduce the quantities
\ba
Y_{3}&=&(p_+ q_+)^2+q_+^2(q_-^2-4 m^2)\,,\\
\label{Y3b}
Y_{4}&=&(p_+ q_+)^2+q_+^2(q_-^2+4 m^2)\,,
\label{Y4b}
\ea
which will be used below in the case of two-photon exchange, as well as the variables
\ba
\varepsilon =\frac{Y_1}{Y_2},\; \varepsilon_1 =\frac{Y_3}{Y_2}, \; \varepsilon_2 =\frac{Y_4}{Y_2}.
\label{eps1}
\ea
relating to the polarization of the virtual photon. In the case $m^2 \to 0$, we have
\ba
\varepsilon=\varepsilon_1=\varepsilon_2=
\frac {(p_+q_+)^2+q_+^2q_-^2}{(p_+ q_+)^2-q_+^2 q_-^2}\, ,
\label{eps12}
\ea
from where it is easy to obtain an expression for the degree of linear polarization
of the virtual photon $\varepsilon$ in the laboratory reference frame, which was defined after Eq. (\ref{Ros}).


\section*{MATRIX ELEMENTS OF THE PROTON CURRENT IN THE TWO-PHOTON
APPROXIMATION}

The matrix elements of the proton current in the two-photon exchange approximation have the form
\ba
\label{MGamu3}
(J_p)^{2\gamma}_{\mu} &=&\overline{u}(q_2) \Gamma^{2\gamma}_{\mu} (q^{2}) u(q_1) \,,
\ea
where $\Gamma^{2\gamma}_{\mu}(q^2)$ can be written in two equivalent representations
\cite{Guichon2003,Brodsky2005,Tjon2007,Borisyuk2007,Borisyuk2008,Borisyuk2011,Kivel2013}
\ba
\label{Gamu4}
\Gamma^{2\gamma}_{\mu}(q^2) = H_{1} \gamma_{\mu} + \frac{H_{2}} {4M}
( \hat q \gamma_{\mu} - \gamma_{\mu} \hat q  )
+ \frac {\hat p_+ (q_+)_{\mu}}{4M^2} H_3 ,~~\\
\Gamma^{2\gamma}_{\mu}(q^2) =H_M \gamma_{\mu}-
\frac{(q_+)_{\mu}}{2M} \, H_2 \, + \frac {\hat p_+
(q_+)_{\mu}}{4M^2}\, H_3\, .~~
\label{Gamu3}
\ea
Here, the complex proton form factors are denoted as $H_M$, $H_1$ $H_2$, and $H_3$
in order to have the direct relation to the standard proton form factors $G_M$, $F_1$, and $F_2$ in
the Born approximation:
\ba
H_1^{1\gamma}=F_1, H_2^{1\gamma}=F_2, H_M^{1\gamma}=G_M, H_3^{1\gamma}=0.
\ea
The matrix elements of the proton current (\ref{MGamu3}) calculated
with Eqs. (\ref{MQED1}), (\ref{P21pp2}), (\ref{P21pm2}), and (\ref{Gamu4}) have the form
\ba
\label {Jepep-pp3}
( J^{ \delta,\delta }_{p} )^{2\gamma}_{\mu} &=& 2 M  \Big(H_1 -\tau_p H_2+\nu H_3\Big)( b_{0} )_{\mu} , \\
\label {Jepep-pm3}
 ( J^{ -\delta,\delta }_{p} )^{2\gamma}_{\mu}&=& -2 \delta M \sqrt{\tau_p} \times\\
& \times& \Big((H_1+H_2) (b_{\delta })_{\mu} +\frac{p_+ b_{\delta}}{4M^2} H_3(q_+)_{\mu}\Big) .\nn 
\ea

The matrix elements of the proton current (\ref{MGamu3}) calculated
with Eqs. (\ref{MQED1}), (\ref{P21pp2}), (\ref{P21pm2}), and (\ref{Gamu3}) have the form
\ba
\label {Jepep-pp2}
( J^{ \delta,\delta }_{p} )^{2\gamma}_{\mu} &=& 2 M  \Big(H_M -\tau_1 H_2+\nu H_3\Big)( b_{0} )_{\mu}, \\
\label {Jepep-pm2}
( J^{ -\delta,\delta }_{p} )^{2\gamma}_{\mu}&=& -2 \delta M \sqrt{\tau_p} \times \\
& \times &\Big(H_{M} (b_{\delta })_{\mu} +\frac{p_+ b_{\delta}}{4M^2} H_3(q_+)_{\mu}\Big) , \nn
\ea
where
\ba
\tau_1=\frac{q_{+}^2}{4M^2}=1+\tau_p,  \tau_p=\frac{Q^2}{4M^2},
\nu=\frac{p_+q_+}{4M^2}=\frac{s-u}{4M^2}. \nn
\ea
Comparing the matrix elements of the proton current (\ref{Jepep-pp3}) and (\ref{Jepep-pp2}),
we obtain the following expression for the "generalized" electric form factor introduced
in \cite{Borisyuk2007,Borisyuk2008,Borisyuk2011}:
\ba
\label{H_E}
&&\CG_E\equiv H_E +\nu H_3, \\
&&H_{E} \equiv H_1 -\tau_p H_2=H_{M}- \tau_1 H_{2}  \,, \\
&&H_M\equiv H_1 + H_2.
\ea

In this case, the form factors $H_E$ and $H_M$ are naturally transformed to $G_E$ and $G_M$
in the case of one-photon exchange:
\ba
\label{H_E2}
H_E^{1\gamma}&=&H_1^{1\gamma} - \tau_p H_2^{1\gamma}= H_M^{1\gamma} -
\tau_1 H_2^{1\gamma}= \nn \\
&=&F_1-\tau_p F_2=G_M - (1+\tau_p)F_2=G_E .\nn
\ea

As a result, the matrix elements of (\ref{Jepep-pp3}) and (\ref{Jepep-pp2}) can
be represented in the form
\ba
\label {Jepep-pp4} (
J^{\delta,\delta  }_{p} )^{2\gamma}_{\mu} = 2 M \CG_E ( b_{0})_{\mu}\, .
\ea

This expression is similar to Eq. (\ref{Jepep-pp0}) in the one-photon approximation and,
therefore, $\CG_E$ can be called the generalized electric form factor of the proton.
In the case of two-photon exchange, the form factor 
\cite{Borisyuk2007,Borisyuk2008}
\ba
\CG_M=H_M+\varepsilon \nu H_3=H_1+H_2+\varepsilon \nu H_3
\ea
serves as the generalized magnetic Sachs form factor.

In terms of $\CG_M$, Eq. (\ref{Jepep-pm2}) has the form
\ba
\label {Jepep-pm5}
( J^{ -\delta,\delta }_{p} )^{2\gamma}_{\mu}&&= -2 \delta M \sqrt{\tau_p}\times  \\
&&\times \Big( (\CG_M - \varepsilon \nu H_3) (b_{\delta })_{\mu}
+\frac{p_+ b_{\delta}}{4M^2} H_3(q_+)_{\mu}\Big). \nn 
\ea

We note that matrix elements of Eqs. (\ref{Jepep-pm2}) and (\ref{Jepep-pm5})
include a factor of $\sqrt{\tau_p}$, which ensures the dominant contribution to the cross
section for transitions of the proton with spin flip at $\tau_p \gg 1$; in the one-photon
approximation, they are transformed to Eq. (\ref{Jepep-pm0}).


\bc
{\bf CROSS SECTION FOR $e \vec p \to e \vec p$  SCATTERING \\IN THE TWO-PHOTON APPROXIMATION}
\ec

Squares of the magnitudes of the amplitudes $|T^{\pm \delta,\delta}|^2$  (\ref{T2pm})
calculated by the general formula (\ref{Tquad1}) in an arbitrary reference frame
using the matrix elements of Eqs. (\ref{Jepep-pp4}) and (\ref{Jepep-pm5}) in the two-photon approximation
in the case of unpolarized electrons have the form
\ba
\label{TPEpp1}
|T^{+ \delta,\delta}|^2&=& \frac{|\CG_E|^2}{(1+\tau_p)}Y_1,\\
|T^{- \delta,\delta}|^2&=&\frac {\tau_p Y_2}{(1+\tau_p)}\Bigl(
|\CG_M|^2 + \varepsilon_1 \varepsilon_2 \tau_p (1+\tau_p) |H_3|^2 \Bigr).~~~~~~
\label{F3quad1}
\ea

If the mass of the electron can be neglected, then
\ba
\varepsilon=\varepsilon_1=\varepsilon_2
 = \frac{\nu^2- \tau_p (1+\tau_p)}{\nu^2+\tau_p( 1+\tau_p) }\,.
\label{eps13}
\ea
As a result, Eq. (\ref{F3quad1}) at $m^2=0$ has the form
\ba
|T^{- \delta,\delta}|^2= \frac{\tau_p \, Y_2}{(1+\tau_p)}\,\Bigl(
|\CG_M|^2 + \varepsilon^2 \,\tau_p (1+\tau_p) \,|H_3|^2 \Bigr).~~
\label{F3quad2}
\ea
We present the useful relation between $\varepsilon, \nu$, and $\tau_p $
\ba
\nu^2 =\tau_p(1+\tau_p)\,\frac{1+\varepsilon}{1-\varepsilon}\,,
\label{nu2gamma}
\ea
which allows reducing the term containing $|H_3|^2$ in Eq. (\ref{F3quad2})
to the form obtained in \cite{Borisyuk2008}
\ba
|T^{- \delta,\delta}|^2= \frac{\tau_p \, Y_2}{(1+\tau_p)}\Bigl(
|\CG_M|^2 + \varepsilon^2 \frac{1-\varepsilon}{1+\varepsilon}\,|\nu H_3|^2 \Bigr).~~~
\label{F3quad3}
\ea
Since the contribution of the amplitude $|H_3|$ vanishes in the Born approximation
and has the smallness order $O(\alpha)$, the last terms in Eqs. (\ref{F3quad1})
and (\ref{F3quad2}) can be neglected. As a result, the cross section in the two-photon
approximation is given by an expression similar to Eq. (\ref{crepep1}),
where the Sachs form factors $G_E$ and $G_M$ are replaced by the generalized form factors
$\CG_E$ and $\CG_M$
\ba
\label{crepepTPE1}
&& ~~~~~~~~~~~~~~~~~~~~~~~~ \frac {d \sigma_{\delta_1, \delta_2}}{d |t|} \\
&& = \frac{\pi \alpha^2 }{4 I^2 (1+\tau_p)} \left(  \omega_+ |\CG_E|^{2}  Y_{1} + \omega_-
\tau_p |\CG_M|^{2}  Y_{2}  \right) \frac{1}{q^4 }. \nn
\ea
In the rest system of the initial proton,
\ba
\label{RosPoltpe2}
\frac{d\sigma_{\delta_1, \delta_2}} {d\Omega_e}= 
\omega_{+} \sigma^{\uparrow\uparrow}+\omega_{-}\sigma^{\downarrow\uparrow},~~~~~~~\\
\sigma^{\uparrow\uparrow}=\sigma_M \, |\CG_E|^2 ,\;\;
\sigma^{\downarrow\uparrow}=\sigma_M \frac{\tau_p}{\varepsilon} \,
|\CG_M|^2\,,
\label{RosPol2tpea}
\ea
where
\ba \sigma_M= \frac{\alpha^2E_2\cos^2(\theta_e/2)}
{4E_1^{\,3}\sin^4(\theta_e/2)} \frac{1}{1+\tau_p}\,.
\ea

In Eq. (\ref{RosPoltpe2}), $\delta_{1}$ and $\delta_{2}$ appearing in $\omega_{\pm}$
are doubled projections of the spins of the initial and final protons
on the common $\vecc c$ axis of spin projections (\ref{LSO}),
so that $-1\leqslant \delta_{1,2}\leqslant 1$.

According to Eq. (\ref{RosPoltpe2}), if electron–proton scattering occurs without
proton spin flip ($\delta_1=1, \delta_2=1$), the contribution to the cross section
comes from only the term containing $|\CG_E|^2$ because the polarization
factors $\omega_{+}$ and $\omega_{-}$ at $|\CG_E|^2$ and $|\CG_M|^2$ are
unity ($\omega_{+}=1$) and zero ($\omega_{-}=0$). If scattering occurs with proton
spin flip ($\delta_1=1, \delta_2=-1$), the contribution to the cross section
comes from only the term containing $|\CG_M|^2$, because the polarization factors
$\omega_{+}$ and $\omega_{-}$ at $|\CG_E|^2$ and $|\CG_M|^2$ are zero ($\omega_{+}=0$)
and unity ($\omega_{-}=1$).

In the case of unpolarized electrons and protons, Eq. (\ref{RosPoltpe2}) gives
the Rosenbluth cross section in the two-photon approximation denoted
as $\sigma_R=d\sigma / d\Omega_e$:
\ba
\label{GRoss}
\sigma_R
=\sigma^{\uparrow\uparrow} + \sigma^{\downarrow\uparrow}\,.
\ea
Consequently, the physical meaning of the separation into two terms in Eq. (\ref{GRoss})
containing only $|\CG_E|^2$ and $|\CG_M|^2$ is the sum of the cross sections for processes without
and with proton spin flip when the initial proton at rest is fully polarized along
the direction of motion of the final proton.

\section*{CONCLUSIONS}

The differential cross section for elastic electron\- proton scattering $e \vec p \to e \vec p$,
where the initial and final protons are polarized and have a common axis of spin
projections has been calculated in an arbitrary reference frame taking into account
two-photon exchange within the phenomenological description of the electromagnetic
electron–proton interactions. The resulting expression (\ref{RosPoltpe2}) for the cross
section in the laboratory reference frame can be used to measure the squares of absolute
values of the generalized Sachs form factors $|\CG_E|^2$ and $|\CG_M|^2$ in processes
without and with proton spin flip in the case where the initial proton at rest
is fully polarized along the direction of the motion of the final proton.

\section*{ ACKNOWLEDGMENTS}
We are grateful to V.S. Fadin for stimulating discussions of the results.

\section*{FUNDING}
R.~E.~ Gerasimov acknowledges the support of the Russian Foundation for Basic Research
(project no. 19-02-00690).


\begin{thebibliography}{60}

\bibitem{Rosen} M.\,N.\, Rosenbluth,
\href{https://doi.org/10.1103/PhysRev.79.615}
{Phys. Rev. {\bf 79}, 615 (1950)}.

\bibitem{Hof58}  R. Hofstadter, F. Bumiller, M. Yearian,  
\href{https://doi.org/10.1103/RevModPhys.30.482}{Rev. Mod. Phys. {\bf 30}, 482 (1958).}

\bibitem{Dombey} N. Dombey,
\href{https://doi.org/10.1103/RevModPhys.41.236}{Rev. Mod. Phys. {\bf 41}, 236 (1969).}

\bibitem{Rekalo74}
A. I. Akhiezer and M. P. Rekalo, Sov. J. Part. Nucl. {\bf 4}, 277 (1974).

\bibitem{AR} 
A. I. Akhiezer and M. P. Rekalo, {\it Electrodynamics of Hadrons} (Naukova Dumka, Kiev, 1977) [in Russian].

\bibitem{GL97}
M. V. Galynskii and M. I. Levchuk, Sov. J. Phys. At. Nucl. {\bf 60} (11), 1855 (1997).

\bibitem{Jones00}
M.K.~Jones, K.A.~Aniol, F.T.~Baker {\em et al.},
\href{https://dx.doi.org/10.1103/PhysRevLett.84.1398}{Phys. Rev. Lett. {\bf 84},  1398 (2000).}

\bibitem{Gay01}
O. Gayou, K. Wijesooriya, A. Afanasev {\em et al.},
\href{https://dx.doi.org/10.1103/PhysRevC.64.038202}{Phys.\ Rev.\ C {\bf 64}, 038202 (2001).}

\bibitem{Gay02} O. Gayou, K.A. Aniol, T. Averett {\em et al.,} 
\href{https://dx.doi.org/10.1103/PhysRevLett.88.092301}{Phys.\ Rev.\ Lett. {\bf 88}, 092301 (2002).}

\bibitem{Pun05} V. Punjabi, C.F. Perdrisat,  K.A. Aniol {\em et al.},
\href{http://dx.doi.org/10.1103/PhysRevC.71.055202}{Phys. Rev. C {\bf 71}, 055202 (2005)};
Erratum-ibid. \href{https://doi.org/10.1103/PhysRevC.71.069902}{Phys. Rev. {\bf C 71}, 069902 (2005).}

\bibitem {Puckett10} A. J. R. Puckett, E. J. Brash, M. K. Jones {\em  et al.},
\href{https://doi.org/10.1103/PhysRevLett.104.242301}{Phys.\ Rev.\ Lett.\ {\bf 104}, 242301 (2010).}

\bibitem {Puckett12}
A.\,J.\,R.\, Puckett, E.\, J.\, Brash, O.\, Gayou, {\em et al.},
\href{https://doi.org/10.1103/PhysRevC.85.045203}{Phys. Rev. C {\bf 85}, 045203 (2012).}

\bibitem{Qattan}  I.\,A.\, Qattan, J.\,Arrington, R.\,E.\,Segel, {\em et al.},
\href{https://doi.org/10.1103/PhysRevLett.94.142301}{Phys. Rev. Lett. {\bf 94},  142301 (2005).}

\bibitem{Perdrisat2007} C. F. Perdrisat, V.\,Punjabi, M.\, Vanderhaeghen,
\href{https://doi.org/10.1016/j.ppnp.2007.05.001}{Prog. Part. Nucl. Phys. {\bf 59}, 694 (2007).}

\bibitem{ETG15}   S.~Pacetti, R.~Baldini Ferroli and E.~Tomasi-Gustafsson,
\href{ https://doi.org/10.1016/j.physrep.2014.09.005}{Phys. Rept. {\bf 550-551}, 1 (2015).}

\bibitem{Guichon2003}
P. A. M. Guichon and M. Vanderhaeghen,
\href{https://doi.org/10.1103/PhysRevLett.91.142303}{Phys. Rev. Lett. {\bf 91}, 142303 (2003).}

\bibitem{Brodsky2005}
A. V. Afanasev, S. J. Brodsky, C. E. Carlson, Y.-Ch. Chen, and M. Vanderhaeghen,
\href{https://doi.org/10.1103/PhysRevD.72.013008}{Phys. Rev. D. {\bf 72}, 013008 (2005).}


\bibitem{Tjon2007} J. Arrington, W.Melnitchouk, and J. A. Tjon,
\href{https://doi.org/10.1103/PhysRevC.76.035205}{ Phys. Rev. C {\bf 76}, 035205 (2007).}

\bibitem{Borisyuk2007} D. Borisyuk and A. Kobushkin,
\href{https://doi.org/10.1103/PhysRevC.75.038202}{Phys. Rev. C {\bf 75}, 038202 (2007).}

\bibitem{Borisyuk2008} D. Borisyuk and A. Kobushkin,
\href{http://doi.org/10.1103/PhysRevC.78.025208}{Phys. Rev. C {\bf 78}, 025208 (2008).}

\bibitem{Borisyuk2011} D. Borisyuk, A. Kobushkin,
\href{https://doi.org/10.1103/PhysRevD.83.057501}{Phys. Rev. D. {\bf 83}, 057501 (2011).}

\bibitem{Kivel2013} N. Kivel and M. Vanderhaeghen,
\href{https://doi.org/10.1007/JHEP04(2013)029}{J. High Energy Phys. {\bf 04}, 029 (2013).}

\bibitem {Meziane11} M. Meziane, J. Brash, R. Gilman {\em et al.}
\href{https://doi.org/10.1103/PhysRevLett.106.132501}{Phys. Rev. Lett. {\bf 106}, 132501 (2011).}

\bibitem {Puckett17} A. J. R. Puckett,  E. J. Brash, M. K. Jones {\em et al.},
\href{https://doi.org/10.1103/PhysRevC.96.055203}{Phys. Rev. C {\bf 96}, 055203 (2017).}

\bibitem{Carlson2007}  C.E. Carlson and M. Vanderhaeghen,
\href{https://doi.org/10.1146/annurev.nucl.57.090506.123116}{Ann. Rev. Nucl. Part. Sci. {\bf 57}, 171 (2007).}

\bibitem{Arrington2011} J. Arrington, P.\,G.\,Blunden, and W.\, Melnitchouk,
\href{https://doi.org/10.1016/j.ppnp.2011.07.003}{Prog. Part. Nucl. Phys. {\bf 66}, 782 (2011).}

\bibitem{Gramolin2015} I. A. Rachek, J. Arrington, V. F. Dmitriev {\em et al.},
\href{https://doi.org/10.1103/PhysRevLett.114.062005}{Phys. Rev. Lett. {\bf 114}, 062005 (2015).}

\bibitem{CLAS2015} D. Adikaram, D. Rimal, L.B. Weinstein {\em et al.}, 
\href{https://doi.org/10.1103/PhysRevLett.114.062003}{Phys. Rev. Lett.  {\bf 114}, 062003 (2015).}


\bibitem{OLYMPUS2017}
B.S. Henderson, L.D. Ice, D. Khaneft {\em et al.}, 
\href{https://doi.org/10.1103/PhysRevLett.118.092501}{Phys. Rev. Lett. {\bf 118}, 092501 (2017).}

\bibitem{JETPL18}
M. V. Galynskii,
\href{https://doi.org/10.1134/S0021364019010089}{JETP Lett. {\bf 109}, 1 (2019).}

\bibitem{Sik84} 
S. M. Sikach, Vesti AN BSSR, Ser. Fiz.-Mat. Navuk {\bf 2}, 84 (1984).

\bibitem{FIF70}  F. I. Fedorov,
\href{https://doi.org/10.1007/BF01038044}{Theor. Math. Phys. {\bf 2}, 248 (1970).}

\bibitem{GL} 
F. I. Fedorov, {\it Lorentz Group} (Nauka, Moscow, 1979)
[in Russian].

\bibitem{GS89} 
M. V. Galynskii, L. F. Zhirkov, S. M. Sikach, and F. I. Fedorov, Sov. Phys. JETP {\bf 68}, 1111 (1989).

\bibitem{GS98} 
M. V. Galynskii and S. M. Sikach,
\href{https://sci-hub.tw/10.1134/1.953087}{Phys. Part. Nucl. {\bf 29}, 469 (1998).}

\bibitem{Cedrik18}  C. Lorc\'e,
\href{https://doi.org/10.1103/PhysRevD.97.016005}{Phys. Rev. D {\bf 97}, 016005 (2018)}.

\bibitem{AB} A. I. Akhiezer and V. B. Berestetskii, {\it Quantum Electrodynamics},
3rd ed. (Nauka, Moscow, 1969; Wiley, New York, 1965).

\bibitem{BLP} V. B. Berestetskii, E. M. Lifshitz, and L. P. Pitaevskii,
{\it Course of Theoretical Physics}, Vol. 4: {\it Quantum Electrodynamics}
(Nauka, Moscow, 1989; Pergamon, Oxford, 1982).

\end{thebibliography}
\end{document}